# Unraveling Enhanced Superconductivity in Single-layer FeSe through Substrate Surface Terminations

Qiang Zou[1], Gi-Yeop Kim[2], Jong-Hoon Kang[3], Basu Dev Oli[1], Zhuozhi Ge[1], Michael Weinert[4], Subhasish Mandal[1], Chang-Beom Eom[3], Si-Young Choi[2], and Lian Li[1,*]

[1]Department of Physics and Astronomy, West Virginia University, Morgantown, WV 26506, USA

[2]Department of Materials Science and Engineering, Pohang University of Science and Technology, Pohang 37673, Republic of Korea

[3]Department of Materials Science and Engineering, University of Wisconsin-Madison, Madison, WI 53706, USA

[4]Department of Physics, University of Wisconsin, Milwaukee, WI 53211, USA

**Abstract:** Single-layer FeSe films grown on (001) SrTiO$_3$ substrates have shown a significant increase in superconducting transition temperature compared to bulk FeSe. Several mechanisms have been proposed to explain such enhancement, including electron doping, interfacial electron-phonon coupling, and strong electron correlations. To pinpoint the primary driver, we grew FeSe films on SrTiO$_3$ substrates with coexisting TiO$_2$ and SrO surface terminations. Scanning tunneling spectroscopy revealed a larger superconducting gap of $\Delta$=17 meV for FeSe on TiO$_2$ compared to $\Delta$=11 meV on SrO. Tunneling spectroscopy also showed a larger work function on SrO, leading to reduced charge transfer, as confirmed by angle-resolved photoemission spectroscopy. Scanning transmission electron microscopy revealed distinctive interfacial atomic-scale structures, with the Se-Fe-Se tetrahedral angle changing from 109.9° on SrO to 105.1° on TiO$_2$. Compared to dynamical mean field theory calculations, these results suggest optimal electron correlations in FeSe/TiO$_2$ for enhancing high-temperature superconductivity.

**Introduction**

The sensitive dependence of two-dimensional (2D) materials on their environment often gives rise to emergent properties. The fusion of 2D materials with correlated complex oxide substrates with designed interfaces thus can provide a robust handle to control their quantum phases via interfacial charge transfer doping, electron-phonon coupling, electron correlations, and light-matter interactions. For materials grown on a substrate, epitaxial constraints can generate 2D structural polymorphs prohibited in bulk crystals. For example, metastable topological insulator 1T'-$WSe_2$ can be grown on $SrTiO_3$ (STO) substrates by molecular beam epitaxy (MBE) [1]. More interestingly, single-layer (SL) FeSe epitaxially grown on STO(001) substrates [2-7] showed a transition temperature nearly eight-fold higher than bulk FeSe [8]. Such enhancement can be further tailored by photoexcitation in the STO substrate after UV light exposure. In addition to the 20% increase in the zero resistance $T_C$, the UV light-induced superconducting state is also non-volatile, which can be rapidly reversed by applying voltage pulses to the STO substrate [9], demonstrating the optical and electrical control of quantum states at designed correlated interfaces of 2D materials and 3D complex oxides.

The mechanisms for the enhancement and control of $T_C$ in SL FeSe/STO are still a subject of intense debate, including electron doping [3,6,10-16] and strong electron-phonon coupling involving the STO Fuchs–Kliewer phonons [17-22]. While the evidence for the electron-phonon coupling, i.e., the appearance of replica bands observed in angle-resolved photoemission spectroscopy (ARPES) [17], has been debated [23], recent studies have highlighted the role of localized phonons at the interfacial $TiO_2$ layer [24-25]. An electron doping of ~0.10-0.12 e/Fe has been considered optimal for achieving high $T_C$ at the $TiO_2$ interface [15]. However, recent work on the FeSe/$FeO_x$ interface has shown that comparable or even higher $T_C$ can be achieved with only 40% of that doping level (0.038 e/Fe) [26-27]. On the other hand, a less discussed and possibly more interesting aspect of $T_C$ enhancement in FeSe/STO is the role of electron correlations. Embedded dynamical mean field theory (eDMFT) calculations have indicated that electron correlations in the SL FeSe/STO are orbital-dependent, with the $d_{xy}$ orbital the most correlated and controlled by the Se-Fe-Se tetrahedral angle and doping from interfacial oxygen vacancies [28].

From an experimental perspective, a more intuitive way to unveil the role of electron doping, electron-phonon coupling, and electron correlations in enhancing $T_C$ is to grow epitaxial SL FeSe on the same substrate with different surface terminations. This eliminates the contributions of bulk phonons, thus allowing independent control of two other key parameters: 1) the surface work function, which determines charge transfer and doping, and 2) the atomic geometry of the FeSe tetrahedron that controls electron correlations.

To this end, we prepared (001) STO substrates with $TiO_2$- and SrO- mixed surface terminations. Using scanning tunneling microscopy/spectroscopy (STM/S), we determined a surface work function surface of 3.8 eV and 4.3 eV on the $TiO_2$ and SrO terminations, and superconducting gap of $\Delta$=17 and 10 meV for FeSe/$TiO_2$ and FeSe/SrO, respectively. Furthermore, STEM reveals distinct interfacial structures on the two terminations, with the average Se–Fe–Se angle being 109.9° for FeSe/SrO and 105.1° for FeSe/$TiO_2$. By comparing these experimental observations with eDMFT calculations, our findings demonstrate that the electron correlations are optimal at the FeSe/$TiO_2$ interface for achieving high $T_C$.

**Results**

**Preparation of STO(001) substrate with $TiO_2$ and SrO termination**. As shown in Fig. 1(a), the STO crystal structure allows two possible terminations with SrO and $TiO_2$ on the (001) surface, separated by steps with the half-unit cell height in the c-axis [29-34]. To prepare STO with both terminations, Nb-doped (0.5% wt) $SrTiO_3$ (001) was annealed at 1040 °C for 30 min in an oxygen atmosphere ($9\times10^{-5}$ Torr) in an ultra-high vacuum (UHV) chamber. Three different measurements confirmed the termination of the resulting STO substrate: 1) *ex-situ* lateral force microscopy (LFM) imaging under ambient conditions, 2) *in-situ* atomic resolution STM imaging, and 3) tip-sample separation-dependent tunnel current (I/Z spectroscopy) measurements of the surface work function in UHV. First, the large-scale LFM image indicates that the as-prepared STO substrate exhibits a step-and-terrace topography with wide and narrow terraces, exhibiting different friction contrasts (Fig. 1(b)). Since SrO is relatively more volatile at high temperatures, it is intuitive to attribute the high contrast terrace to the $TiO_2$ termination, consistent with earlier LFM measurements where the $TiO_2$-termination exhibited higher friction than SrO [32].

Such assignment of the surface termination is confirmed by *in-situ* STM imaging, which also shows similar topography with alternating wide and narrow terraces separated by steps with the half-unit cell height of STO (Figs. 1(c-e)). On the wide terrace, atomic-resolved STM imaging reveals a ($\sqrt{13} \times \sqrt{13}$) reconstruction, consisting of a network of truncated octahedra $TiO_5$ (Fig. 1(f)), typical for the $TiO_2$ termination [34]. In contrast, the narrow terrace shows a row structure with a periodicity of ~1.2 nm, three times the in-plane STO lattice constant $a_{STO}$ (Fig. 1(g)). Since no apparent periodicity is observed along the row, the structure is denoted (3×1), which has been observed on the SrO terminated STO(0001) in earlier work [29-30,34].

Finally, the surface work function was determined by fitting the I/Z spectra [Supplementary Fig. S1], which yields a barrier height of 3.8 ± 0.3 eV and 4.7 ± 0.3 eV for the $TiO_2$- and SrO-termination, respectively. The lowered surface work function of $TiO_2$ termination is expected to facilitate higher charge transfer to FeSe from the STO substrate [15-16].

**Superconducting properties of single-layer FeSe films on $TiO_2$ and SrO terminations.** Single-layer FeSe was grown on the mixed-terminated STO(001) substrate at 350 °C followed by annealing at 510 °C for 1-3 hours. STM imaging shows the FeSe films follow the STO substrate's step-terrace topography (Fig. 2(a)). On the same terrace, two domains of different contrasts and widths, separated by a boundary marked by the white dashed line, can be identified as FeSe/$TiO_2$ and FeSe/SrO based on atomic resolution imaging and differential conductance dI/dV measurements. The $TiO_2$ surface is fully covered by FeSe, while for FeSe/SrO, pits are formed due to the desorption of FeSe. The islands near the step edge with the highest contrast is the second layer FeSe, which is not superconducting [36]. The formation of pits on the FeSe/SrO suggests a slightly weaker bonding between the layers. As shown in Supplementary Fig. S3(a), for samples annealed for a longer time, more than 90% of SL FeSe has desorbed on the SrO termination, while the second layer of FeSe remains on the $TiO_2$ termination. This indicates the coupling between FeSe and SrO surface is weaker than between adjacent FeSe layers. This is further confirmed by the line profile, as shown in Fig. S3(b), indicating the height of SL FeSe on the SrO termination is ~8.7 Å. Given the height of $TiO_2$ is 1.7 Å (Fig. 1(d)), the height of SL FeSe on the $TiO_2$ termination is found to be 6.1 Å, indicating stronger coupling.

The atomic STM imaging confirms the difference between FeSe/$TiO_2$ and FeSe/SrO. While a square lattice is seen on both terminations, for SeFe/SrO, an additional stripe modulation is superimposed on the (1×1) lattice along the crystallographic a-axis. Fast Fourier Transform (FFT) analysis indicates the modulation has a (3×1) periodicity. Interestingly, images acquired at energies below the Fermi level still exhibit the same periodicity and similar contrast as those acquired above $E_F$ (Fig. 2(d)). These observations suggest that the initial (3x1) surface reconstruction of the SrO termination remains at the interface, likely due to the weaker interaction with the FeSe layer on top, as discussed above. For the $TiO_2$ termination, on the other hand, the original (√13x√13) structure is no longer visible in STM imaging, consistent with the stronger coupling with the FeSe layer.

The difference between the SL FeSe films on the two terminations is also revealed by comparing the dI/dV tunneling spectra, as shown in Fig. 2(e), where the pronounced valence band peak is at -170 ± 20 meV for FeSe/SrO and is shifted downwards to ∼ -270 ± 20 meV for FeSe/$TiO_2$. As the peak position is associated with the degree of charge doping [14], the shift indicates a higher number of electrons in FeSe/$TiO_2$ consistent with $TiO_2$'s lower work function that favors electron transfer from the STO substrate.

Next, to probe superconductivity, dI/dV spectra are obtained near the Fermi level, as shown in Fig. 2(f) (additional spectra in Supplementary Fig. S2). The spectra are distinct U-shaped with two pairs of coherence peaks, indicating the FeSe films are superconducting on both terminations. However, the superconductivity gap, defined as the separation between peaks, differs; for FeSe/$TiO_2$, the gaps are 17.0 ± 0.5 eV and 11.0 ± 0.5 meV, similar to those reported in earlier studies [2]. For FeSe/SrO, they are 10.5 ± 0.5 and 5.8 ± 0.5meV, significantly lower, indicating a direct connection between the doping level and the superconducting gap.

**ARPES measurements of single-layer FeSe films:** To quantitatively determine the electron doping, we compared the electronic structure of the SL FeSe/STO on full $TiO_2$- and mixed $TiO_2$/SrO-terminations. Figure 3(a) is an STM topographic image of SL FeSe/$TiO_2$/STO substrate after annealing at 510 °C. Note that annealing is typically necessary to achieve superconductivity in SL FeSe/STO films, often resulting in some FeSe desorption appearing as dark pits and white adsorbates (likely Fe clusters). Figures 3(b-c) show the Fermi surface at M pocket and band structure around M and Γ, consistent with earlier work on superconducting FeSe/STO [6,17]. For FeSe on mixed $TiO_2$/SrO-termination (Fig. 3(d)), the Fermi surface at M has the same elliptical shape, however, with a small electron pocket. From the size of electron pockets, the electron density is estimated to be (0.05 ± 0.05) e⁻ per Fe for FeSe on mixed SrO/TiO$_2$ termination, smaller than the (0.10 ± 0.05) e⁻ per Fe on the TiO$_2$ termination. This is further confirmed by the hole bands around the Γ point. For the FeSe/(SrO&TiO$_2$), the top of the hole band is 60 meV below the Fermi level (Fig. 3f), which is 20 meV closer to the Fermi level than that of FeSe/TiO$_2$ (Fig. 3c). Overall, ARPES measurements reveal similar FeSe electronic band structures for both cases but with 50% reduction in electron doping on the mixed SrO&TiO$_2$ termination, consistent with the I-Z spectroscopy results that show larger surface work function for the SrO termination, unfavorable charge transfer.

**Scanning-transmission-electron-microscopy analysis on single-layer FeSe/STO**: To determine the interfacial atomic structures, STEM measurements were carried out on superconducting SL FeSe grown on mixed-terminated STO substrates, capped by a 20-layer FeTe (Supplementary Figs. S4-5). Figures 4(a-b) show the high-angle annular dark field (HAADF) cross-sectional view of the FeSe/SrO and FeSe/TiO$_2$ interfaces, overlaid with atomic structures of FeTe capping layer, SL FeSe, and STO substrate. A significant difference between the two interfaces is an additional layer between FeSe and double TiO$_2$ layers in FeSe/TiO$_2$. The nature of this additional layer is controversial, attributed to extra Se at the interface in some earlier STEM studies [36-37] due to energetically favorable Se O-vacancy complex [38], and not observed in other studies [39]. Our energy dispersive spectroscopy (EDS) analysis indicates that this additional layer is indeed Se, labeled as Se$_3$ in Fig. 4(d). Details of the EDS atomic profiles of the K and L edges of Fe, Se, Ti, and Sr are shown in Fig. S6, which verify that the FeSe/TiO$_2$ interface contains double TiO$_2$ (Ti$_1$ and Ti$_2$) and additional Se (Se$_3$) layers.

In addition, a close examination of the HAADF-filter images of the interfaces reveals that the average Se-Fe-Se angle is 105.1° and 109.9° for the FeSe/TiO$_2$ and FeSe/SrO interfaces, respectively. Those STEM results confirm distinctive interfacial atomic-scale structures and FeSe tetrahedral geometry between SL FeSe on TiO$_2$ and SrO.

**Discussions**

The epitaxial growth on TiO$_2$- and SrO-terminated STO modifies the SL FeSe properties in two significant ways: 1) it changes the surface work function that controls charge transfer and electron doping, 2) it causes distortion of the FeSe tetrahedron and changes the Fe-Se-Fe bond angle that in turn, controls electron correlations [28]. The larger pairing gap on TiO$_2$ termination is consistent with its smaller work function, which favors charge transfer [15-16]. However, recent work on the FeSe/FeO$_x$ interface has shown that higher T$_C$ can be achieved with a much lower doping level [26-27]. This indicates that higher electron doping is not the most critical contributor to enhanced superconductivity. The bulk STO phonon contributions to the electron-phonon coupling are the same for both terminations, thus not responsible for the difference in pairing gaps. Nevertheless, the localized interfacial phonon may still play a significant role, as revealed by recent studies of FeSe/TiO$_2$ termination [24-25]. Similar studies are called for to probe the role of localized phonons at the FeSe/SrO interface.

The trend of higher Tc on the TiO$_2$ termination accompanied by smaller Se-Fe-Se tetrahedral angle is similar to that found in bulk FeSe, where the angle, thus the T$_C$, is also tunable by applied pressure and intercalation of cationic spacer layers. At ambient pressure, the T$_C$ of bulk FeSe is 8 K [8] with a superconducting gap of $\Delta = 2.2$ meV [40] and a Se–Fe–Se angle of ~104º [41]. At an applied pressure of 7 GPa, the Tc increases to 37 K, with a decreased bond angle of ~102° [42-43]. Similarly, the intercalation of a Li$_x$(NH$_2$)$_y$(NH$_3$)$_{1-y}$ spacer layer increases the T$_C$ to 43 K, with increased interlayer spacing and Se–Fe–Se bond angle of 102.9º [44]. These observations are consistent with recent quasiparticle self-consistent GW and dynamical mean field calculations, which showed that a decreased bond angle reduces Se-Fe-Se hoping, and the increased c-axis reduces electronic screening, both of which enhance electron correlations, thus enhancing T$_C$ [45-46].

To examine the role of electron correlations in enhancing T$_C$ in SL FeSe/STO, we carried out eDMFT calculations on SL FeSe/TiO$_2$ and FeSe/SrO (details shown in Supplementary Note and Fig. S7). The results reveal a similar connection between the Se-Fe-Se bond angle and electron correlations, which is optimal for the FeSe/double TiO$_2$/STO, corresponding to the larger superconducting gap.

In conclusion, the epitaxial growth of single-layer FeSe on (001) SrTiO$_3$ substrates with coexisting TiO$_2$ and SrO terminations provides a unique platform to study the key factors driving high-temperature superconductivity. We observed a 60% larger superconducting gap for FeSe/TiO$_2$, which exhibits a distinct interfacial structure with a smaller average Se–Fe–Se angle and larger electron correlations. By independently controlling the surface work function and charge doping, as well as the atomic geometry of the FeSe tetrahedron, our results show that electron correlation is optimal for FeSe/TiO$_2$ in achieving high-temperature superconductivity.

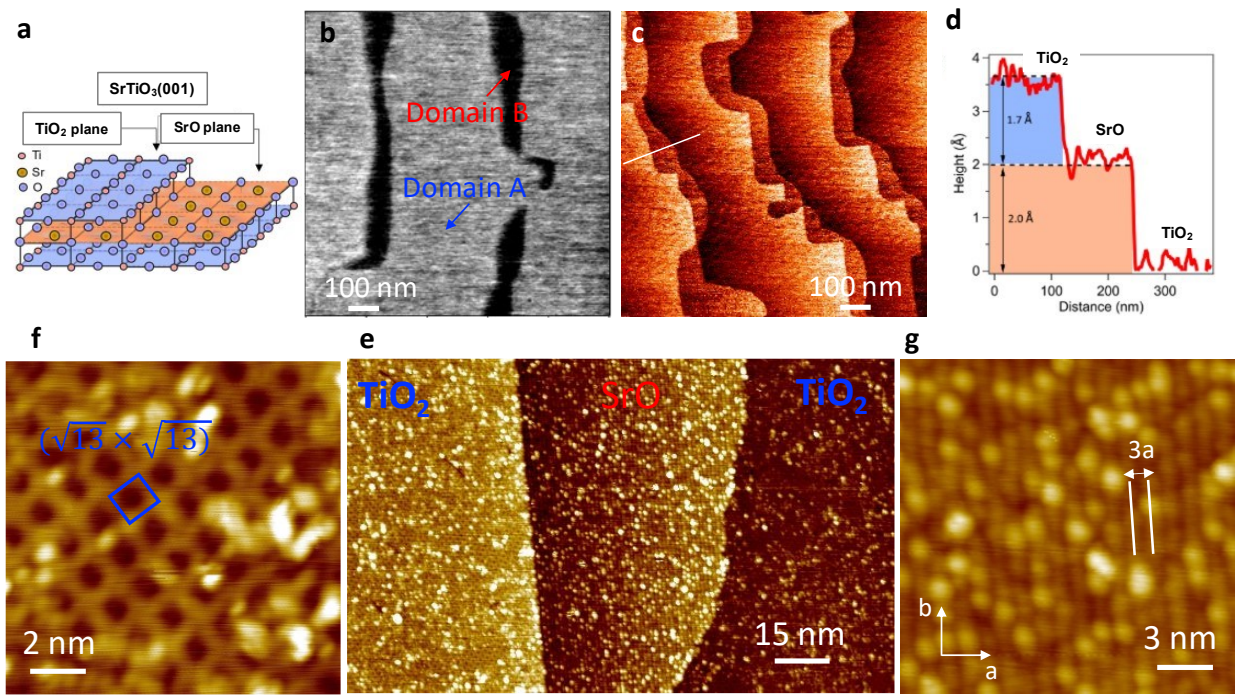

**Figure 1 | SrTiO₃ (001) with mixed SrO and TiO terminations.** **a** Schematic of SrTiO$_3$ substrate with both TiO$_2$ and SrO terminations. LEF (**b**) and STM (**c**) image of the STO(001) substrate with mixed TiO$_2$ and SrO termination ($V_s$= 2.7 V, $I_t$= 0.1 nA). **d** Line profile along the white dashed line in (**c**). **e** STM image of the STO(001) surface with both TiO$_2$ and SrO terminations. **f**, STM image of the TiO$_2$ surface with the ($\sqrt{13} \times \sqrt{13}$) reconstruction ($V_s$= 0.5 V, $I_t$= 0.1 nA). **g** STM image of the SrO surface with the (3x1) reconstruction ($V_s$= 2.0 V, $I_t$= 0.1 nA).

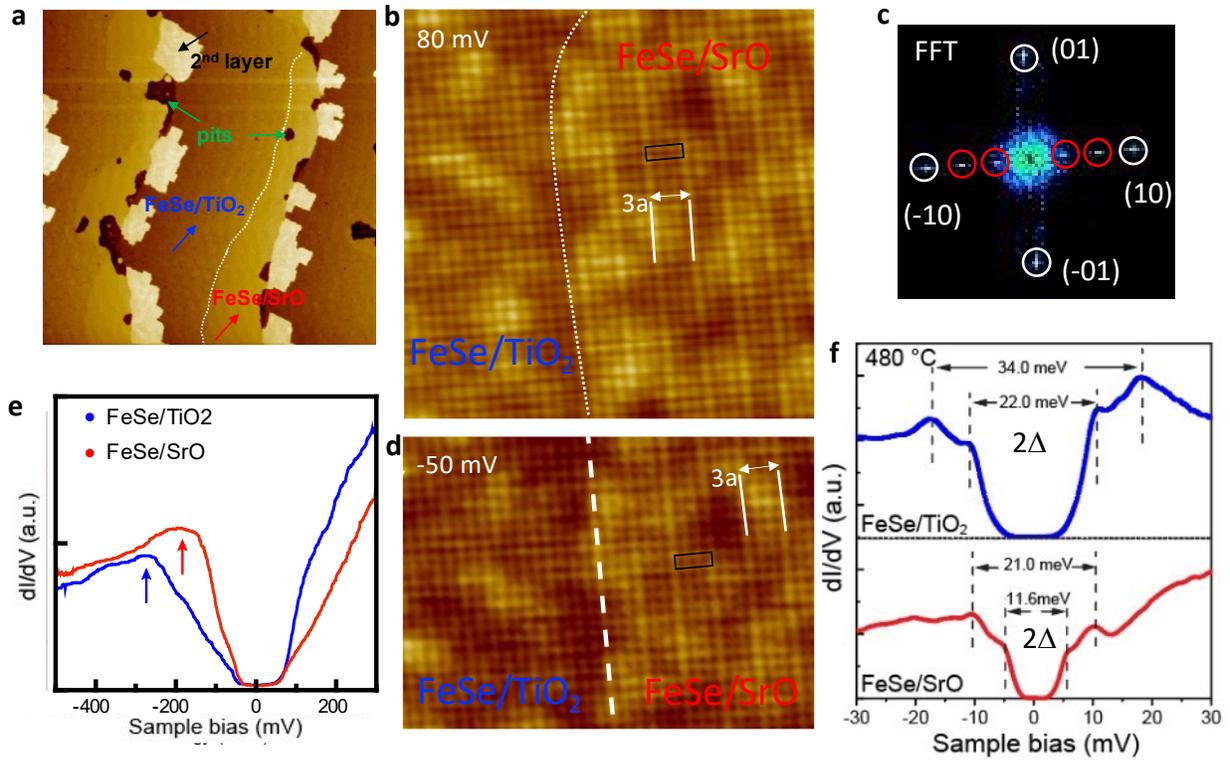

**Figure 2 | Superconducting properties of single-layer FeSe/SrO and FeSe/TiO$_2$. a** large-scale STM image of a single-layer FeSe/TiO$_2$ and FeSe/SrO (V$_s$= 2.7 V, I$_t$= 0.1 nA). **b** Atomic resolution STM image of single-layer FeSe/TiO$_2$ and FeSe/SrO (V$_s$= 50 mV, I$_t$= 0.5 nA). The (3x1) unit cell is marked **c** FFT analysis of the image in (**b**). STM image of the single-layer FeSe/TiO$_2$ and FeSe/SrO taken at -50 meV. The rectangle marks the (3x1) unit cell. **f** Representative dI/dV spectra of FeSe/TiO$_2$ and FeSe/SrO near the Fermi level. The arrows mark the superconducting coherence peaks.

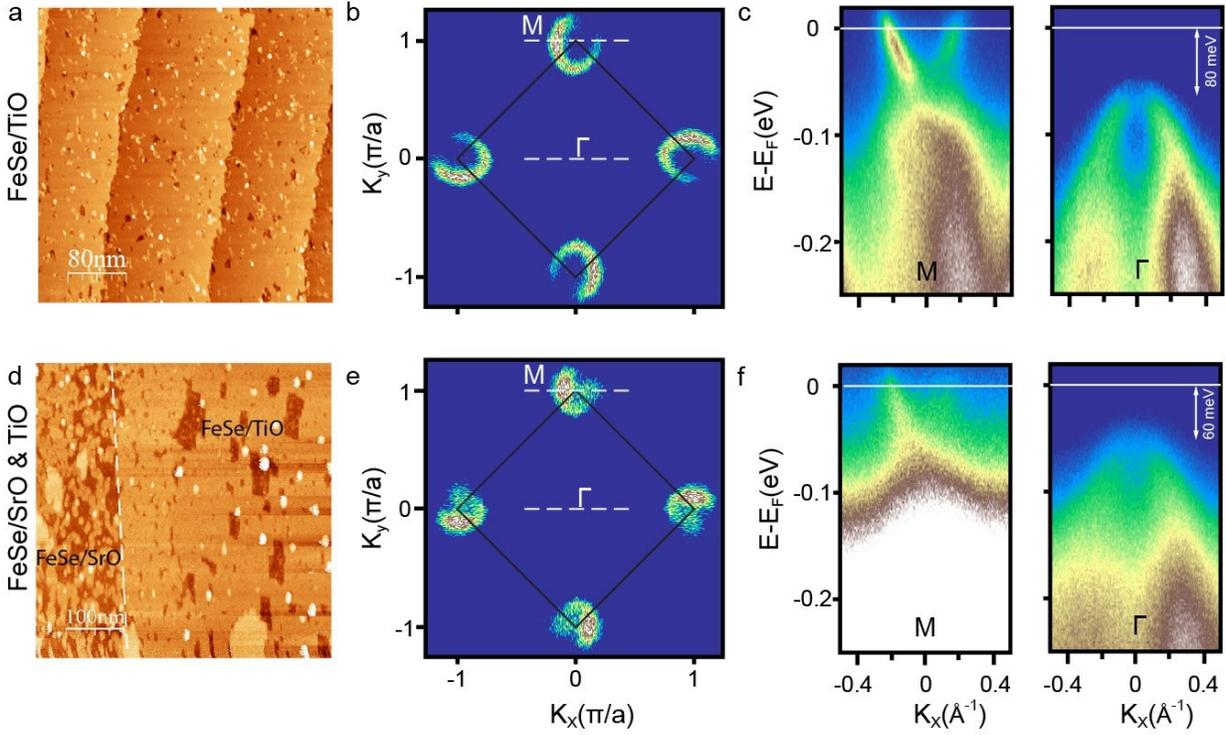

**Figure 3 | ARPES measurements of single-layer FeSe films on the single TiO$_2$-terminated and mixed-terminated STO. a-c** STM topographic image, Fermi surface mapping, and band structure around M and Γ of single-layer FeSe on STO substrate with single TiO$_2$-termiation. **d-f** Same as **a-c**, but on the insulating STO substrate with mixed SrO and TiO$_2$ termination. All measurements were conducted at 80 K.

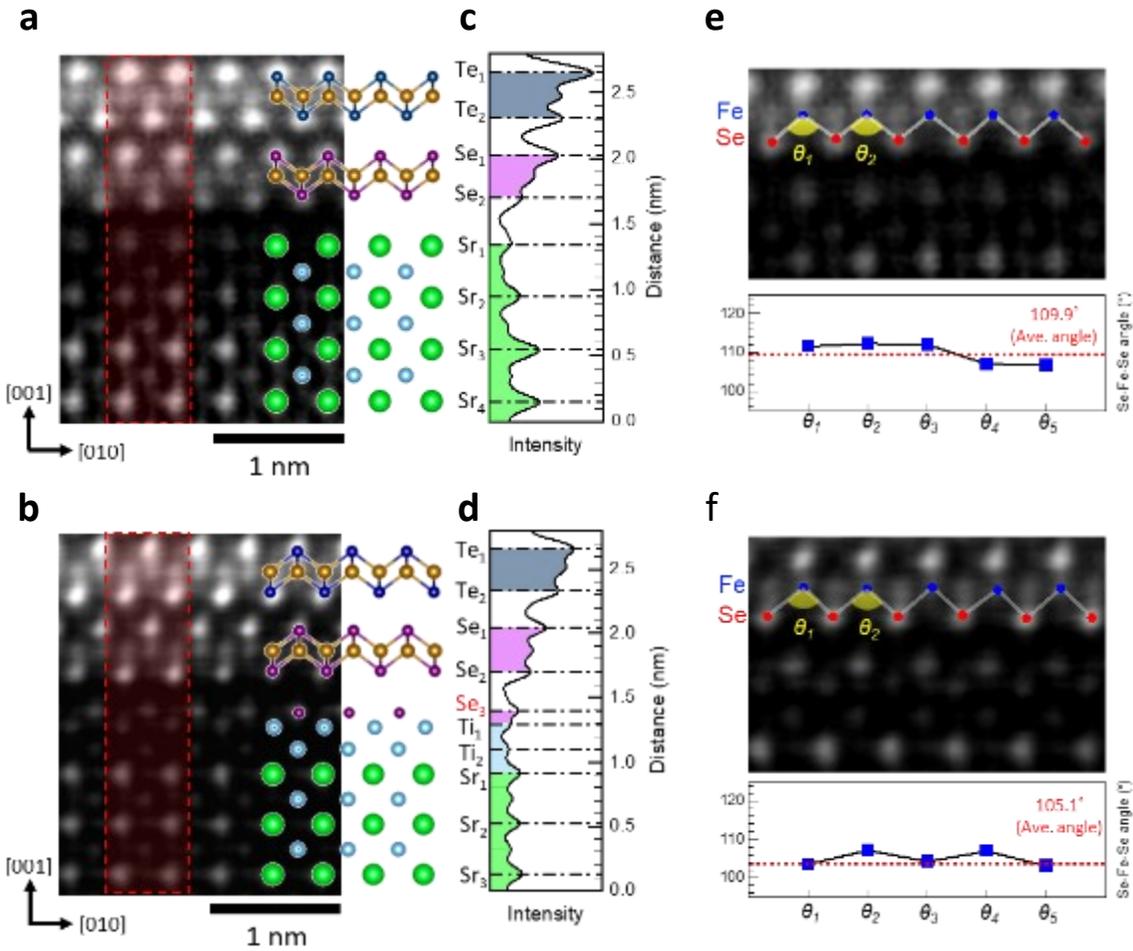

**Figure 4 | STEM imaging of single-layer FeSe on SrO- and TiO$_2$-terminated STO substrates**. **a, b** Cross-sectional HAADF-STEM images show the interface structures of FeSe/SrO and FeSe/TiO$_2$ on two different termination layers of SrTiO$_3$, respectively. **c, d** Intensity profiles along [001] from the areas indicated by dotted boxes in (a) and (b). **e, f** The measured Se-Fe-Se buckling angles of FeSe on the two termination layers of SrO and TiO$_2$ interfaces, respectively. The buckling angle of FeSe/TiO$_2$ is smaller by ~ 5 degrees than that of FeSe/SrO.

## ASSOCIATED CONTENT

**Supporting Information:** methods, supplementary note on eDMFT calculations, dI/dz spectra on $TiO_2$ and SrO termination, spatially resolved dI/dV spectra near Fermi level on FeSe/$TiO_2$ and FeSe/SrO termination, line-profile across a boundary between FeSe/$TiO_2$ and FeSe/SrO, STM image of the FeTe capping layer, TEM sample preparation, STEM-EDS analysis of single-layer FeSe on a $TiO_2$- terminated STO substrate, eDMFT calculated atomic and band structures of single-layer FeSe on $TiO_2$- and SrO-termination STO

## AUTHOR INFORMATION

### Corresponding Author

**Lian Li** − Department of Physics and Astronomy, West Virginia University, Morgantown, West Virginia 26506, United States
Phone: (+1) 304-293-4270; E-mail: lian.li@mail.wvu.edu

### Authors

**Qiang Zou** − Department of Physics and Astronomy, West Virginia University, Morgantown, West Virginia 26506, United States

**Gi-Yeop Kim** − Department of Materials Science and Engineering, Pohang University of Science and Technology, Pohang 37673, Republic of Korea

**Jong-Hoon Kang** − Department of Materials Science and Engineering, University of Wisconsin-Madison, Madison, WI 53706, USA

**Basu Dev Oli** − Department of Physics and Astronomy, West Virginia University, Morgantown, West Virginia 26506, United States

**Zhuozhi Ge** − Department of Physics and Astronomy, West Virginia University, Morgantown, West Virginia 26506, United States

**Michael Weinert** − Department of Physics, University of Wisconsin, Milwaukee, WI 53201, USA

**Subhasish Mandal** − Department of Physics and Astronomy, West Virginia University, Morgantown, West Virginia 26506, United States

**Si-Young Choi** − Department of Materials Science and Engineering, Pohang University of Science and Technology, Pohang 37673, Republic of Kore

**Chang-Beom Eom** − Department of Materials Science and Engineering, University of Wisconsin-Madison, Madison, WI 53706, USA

**Author contributions**

L.L., J.H.K. and C.B.E conceived the idea. J.H.K. and QZ prepared the substrates. C.B.E. L.L. and S.Y.C. supervised the project. G.K. performed the STEM measurements. Q.Z., B.D.O, and Z.Ge. performed the MBE growth and STM/S measurements. S.B. and M.W. performed the calculations. All authors analyzed the data, and Q.Z. and L.L. wrote the paper.

**Notes**

The authors declare no competing financial interest.

## ACKNOWLEDGMENTS

MW and LL acknowledge support from the U.S. Department of Energy, Office of Basic Energy Sciences, Division of Materials Sciences and Engineering under Award No. DE-SC0017632. Work at the University of Wisconsin-Madison was supported by the U.S. Department of Energy (DOE), Office of Science, Basic Energy Sciences (BES) under Award number DE-FG02-06ER46327 (CBE).